\documentclass{article}

\usepackage{arxiv}

\usepackage[utf8]{inputenc}
\usepackage[T1]{fontenc}
\usepackage{url}
\usepackage{booktabs}
\usepackage{amsfonts}
\usepackage{nicefrac}
\usepackage{microtype}
\usepackage{lipsum}
\usepackage{amsmath}
\usepackage{graphicx}
\usepackage{siunitx}
\usepackage[symbol]{footmisc}

\graphicspath{ {images/} }

\title{Transferable and extensible machine learning derived atomic charges for modeling metal-organic frameworks}

\begin{document}

\begin{center}
 {\LARGE Transferable and extensible machine learning derived atomic charges for modeling hybrid nanoporous materials}
\end{center}

\begin{center}
 Vadim Korolev\footnote[1]{\textit{Email address}: \texttt{korolev@colloid.chem.msu.ru}}\textsuperscript{1,2}, Artem Mitrofanov\textsuperscript{1,2}, Ekaterina Marchenko\textsuperscript{3,4}, \\ Nickolay Eremin\textsuperscript{4}, Valery Tkachenko\textsuperscript{1}, Stepan Kalmykov\textsuperscript{2}
\end{center}

\begin{center}
 \textsuperscript{1}\textit{Science Data Software, LLC, 14909 Forest Landing Circle, Rockville, Maryland 20850, United States}\\
 \textsuperscript{2}\textit{Department of Chemistry, Lomonosov Moscow State University, 119991 Moscow, Russia}\\
 \textsuperscript{3}\textit{Laboratory of New Materials for Solar Energetics, Department of Materials Science, Lomonosov Moscow State University, 119991 Moscow, Russia}\\
 \textsuperscript{4}\textit{Department of Geology, Lomonosov Moscow State University, 119991 Moscow, Russia}
\end{center}

\vspace{3mm}

\begin{abstract}
Nanoporous materials have attracted significant interest as an emerging platform for adsorption-related applications. The high-throughput computational screening became a standard technique to access the performance of thousands of candidates, but its accuracy is highly dependent on a partial charge assignment method. In this study, we propose a machine learning model that can reconcile the benefits of two main approaches—the high accuracy of density-derived electrostatic and chemical charge (DDEC) method and the scalability of charge equilibration (Qeq) method. The mean absolute deviation of predicted partial charges from the original DDEC counterparts archive an excellent level of 0.01 e. The model, initially designed for metal-organic frameworks, is also capable of assigning charges to another class of nanoporous materials, covalent organic frameworks, with acceptable accuracy. Adsorption properties of carbon dioxide, calculated by means of machine learning derived charges, are consistent with the reference data obtained with DDEC charges.
\end{abstract}

% keywords can be removed

\section{Introduction}
Metal-organic frameworks (MOFs) form a relatively new class of crystalline porous materials which should be considered a part of the “nanoporous materials genome\cite{boyd2017computational,dang2017nanomaterials}.” Their internal pores, the size of small molecules, lead to confinement effects that strongly affect adsorbates. The resulting properties may be useful for multiple applications, such as catalysis\cite{liu2014applications,dhakshinamoorthy20192d,xu2019functional,yang2019catalysis,pascanu2019metal}, gas storage/separation\cite{li2012metal,ma2010gas,yu2017co2,trickett2017chemistry,jiang2019computational}, and especially in clean energy-related fields\cite{li2013metal,xia2015metal,wang2016metal,zhao2016metal,wang2017metal,zhang2017metal,liang2018pristine}. Due to the enormous structural diversity, the number of experimentally synthesized MOFs has now reached several thousand\cite{chung2019advances}. Computationally developed MOF databases contain up to hundreds of thousands of structures\cite{wilmer2012large}. Thus, providing synthesis, characterization, and testing of all promising candidates for a desired absorbance-related application is impossible from a practical perspective. It is not surprising that \textit{in silico} design has become the main approach to provide large-scale materials screening studies in the field\cite{boyd2017computational}.

Electronic structure calculations based on the density functional theory (DFT) provide an appropriate accuracy in matching experimentally measured data such as adsorbate interaction energies\cite{grajciar2011understanding}. However, similar to experimental studies, large-scale materials screening studies with DFT calculations are disappointingly time-consuming, and they can be carried out only for a limited set of structures. From a more general perspective, DFT simulations are challenging for any subclass of nanoporous materials\cite{coudert2013systematic,fischer2016benchmarking}. In most cases, calculations which use classical force fields provide an optimal trade-off between accuracy and computational cost, especially when screening a large number of MOFs. A starting point and a “bottleneck” of force field calculations is the atomic charge assignment required to describe the Coulomb contribution to interatomic potential. Semiempirical methods such as the charge equilibration\cite{rappe1991charge} (QEq), the extended charge equilibration\cite{wilmer2011towards,wilmer2012extended} (EQEq), and the periodic charge equilibration\cite{ramachandran1996toward} (PQeq) do not require direct electronic structure calculations, and they are strongly preferable under computational cost aspect. However, these methods enormously overestimate the values of point charges for some elements, which may eventually lead to unreliable computational results\cite{kadantsev2013fast}. Methods of charge assignment based on partitioning of the electron density, such as ChelpG\cite{breneman1990determining}, density-derived electrostatic and chemical\cite{manz2010chemically} (DDEC), and repeating electrostatic potential extracted atomic\cite{campana2009electrostatic} (REPEAT) have been applied only for relatively small sets\cite{nazarian2016comprehensive} of structures due to their high computational cost.

\begin{figure}[h]
  \centering
  \includegraphics[height=8cm]{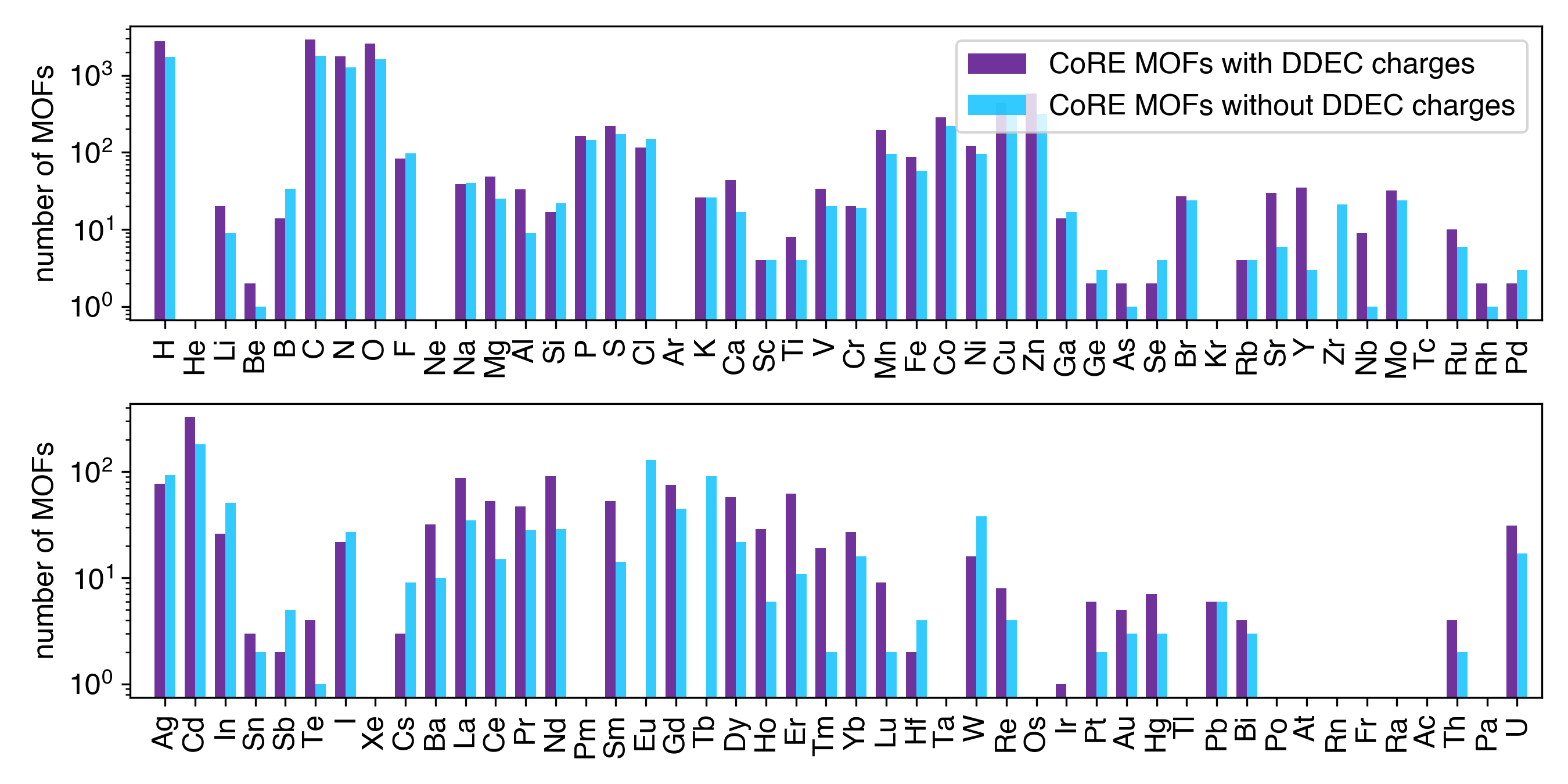}
  \caption{The frequency of elements in the set of structures from the CoRE MOF 2014 database.}
  \label{fig:fig_1}
\end{figure}

Thus, highly parameterized \textit{ad hoc} methods and \textit{ab initio} calculations provide two main options in terms of the balance between accuracy and time/computational cost restraints\cite{manz2010chemically,campana2009electrostatic,manz2012improved}. Unfortunately, both face a substantial issue—every new structure requires distinct calculations and charge assignment. Even with advanced QEq parameters, known as MOF electrostatic-potential-optimized (MEPO) QEq parameters\cite{kadantsev2013fast}, highly parameterized methods do not allow charge assignment for structures that are significantly distinct from ones used in the fitting procedure\cite{ongari2018evaluating}. Xu and Zhong\cite{xu2010general} proposed a calculation-free method based on the assumption that the atoms with the same bonding connectivity have identical charges. Unfortunately, the connectivity-based atom contribution (CBAC) method has an inherent limitation—it can be applied only for atoms with the same types of local environments presented in the training set.

The abovementioned challenges also arise for the molecular charge assignment. However, an alternative approach, unrelated to “classical” computational techniques, has been introduced recently. It has been shown that machine learning (ML) and particularly deep learning techniques can reconcile quantum mechanics accuracy and low computational cost for partial charge calculations\cite{nebgen2018transferable,bleiziffer2018machine,sifain2018discovering}. Moreover, ML models can provide \textit{transferability} and \textit{extensibility} of charge prediction. In other words, it is possible to assign partial charges in systems that differ significantly from the ones that were used to implement ML models. Extensibility usually relates to the system size (number of atoms in a unit cell in the case of periodic systems). Transferability can be interpreted in a broader sense, and it usually involves chemical diversity.

Taking into account that the number of experimentally/computationally characterized MOFs is several orders of magnitude less than the corresponding number of molecules, transferable and extensible charge assignment for MOFs is a complex task, at least because of data scarcity. To overcome this issue, we implemented high-interpretable ML models on the most numerous set of DDEC charges available. We demonstrated the extensibility of the proposed approach by generating self-consistent, i.e., nearly electroneutral, sets of partial charges for Computation-Ready Experimental (CoRE) MOF structures that contain up to ten thousand atoms in a unit cell. The transferability of the implemented models was demonstrated on the example of another emergent subclass of nanoporous materials genome—covalent organic frameworks (COFs).

\newpage
\section{Methods}
\subsection{Reference DFT simulations}

Recent studies have shown that the availability of materials data strongly affects the predictive capability of ML models\cite{faber2016machine,schmidt2017predicting,zhang2018strategy}. Unfortunately, materials datasets are typically far less numerous than the molecular ones; in the worst-case scenario, the number of samples is in the dozens\cite{zhang2018strategy,kumar2018machine,yamada2019predicting}. High computational cost for generating new data points usually leads to data scarcity; this is particularly true for partial charge assignment with high-precision ab initio calculations. Moreover, such data diversity can restrict the predictive capability of the ML model in unexplored domains (outside its applicability domain); in other words, it can limit the ML model’s transferability.

The most reliable set of partial charges for MOFs using DFT calculations and the DDEC charge partitioning approach was presented by Nazarian et al\cite{nazarian2016comprehensive}. Atomic point charge assignment was provided for 2932 experimentally synthesized MOF structures, most of them were collected from CoRE MOF database (version 2014)\cite{chung2014computation}. Solvent molecules, typically trapped in experimentally resolved MOFs, and atoms with partial occupancy/symmetry-related copies of atoms were removed, in order to prepare MOFs for molecular simulations. For the same reason, highly disordered structures were also excluded from consideration. Charge assignment was provided using plane-wave DFT calculations and the DDEC charge partitioning method. Optimization of the geometry of structures was not employed due to its negligible influence on the resulting point charges. In particular, high-quality charges generated by the DDEC method are designed to reproduce electrostatic interactions even outside the van der Waals radius of involved atoms. This is especially crucial for adsorption-related simulations.

It should be noted that the total number of distinct materials in the CoRE MOF 2014 database (3852) is significantly larger than the number of structures for which the electron density was successfully computed (2932). The divergence in calculations occurred mainly due to the large size of a unit cell of corresponding structures and concomitant difficulties such as virtual memory limit exceedance and issues with k-point grid density requirements. In addition, due to computational difficulties, the presence of \textit{f}-elements is more prominent in the unlabeled part of the database. Thus, there are no DDEC charges for Eu and Tb MOFs. Whereas lanthanide-containing MOFs are highly attractive due to their luminescent properties\cite{rocha2011luminescent,cui2014lanthanide}. Therefore, here we define the possibility to model structures containing rare-earth elements as an essential characteristic of a useful ML algorithm. The abovementioned subset of 2932 MOFs was used to probe the extensibility of the presented ML models trained on structures from the CoRE MOF database with relatively small unit cells and lack of \textit{f}-elements. The total number of samples (lattice sites and corresponding partial atomic charges) in the training dataset is approximately four hundred thousand, so it does not seem diverse or sparse. However, the training dataset is considered significantly diverse in a structural and chemical sense (Figure \ref{fig:fig_1}).

\begin{figure}[h]
  \centering
  \includegraphics[height=7cm]{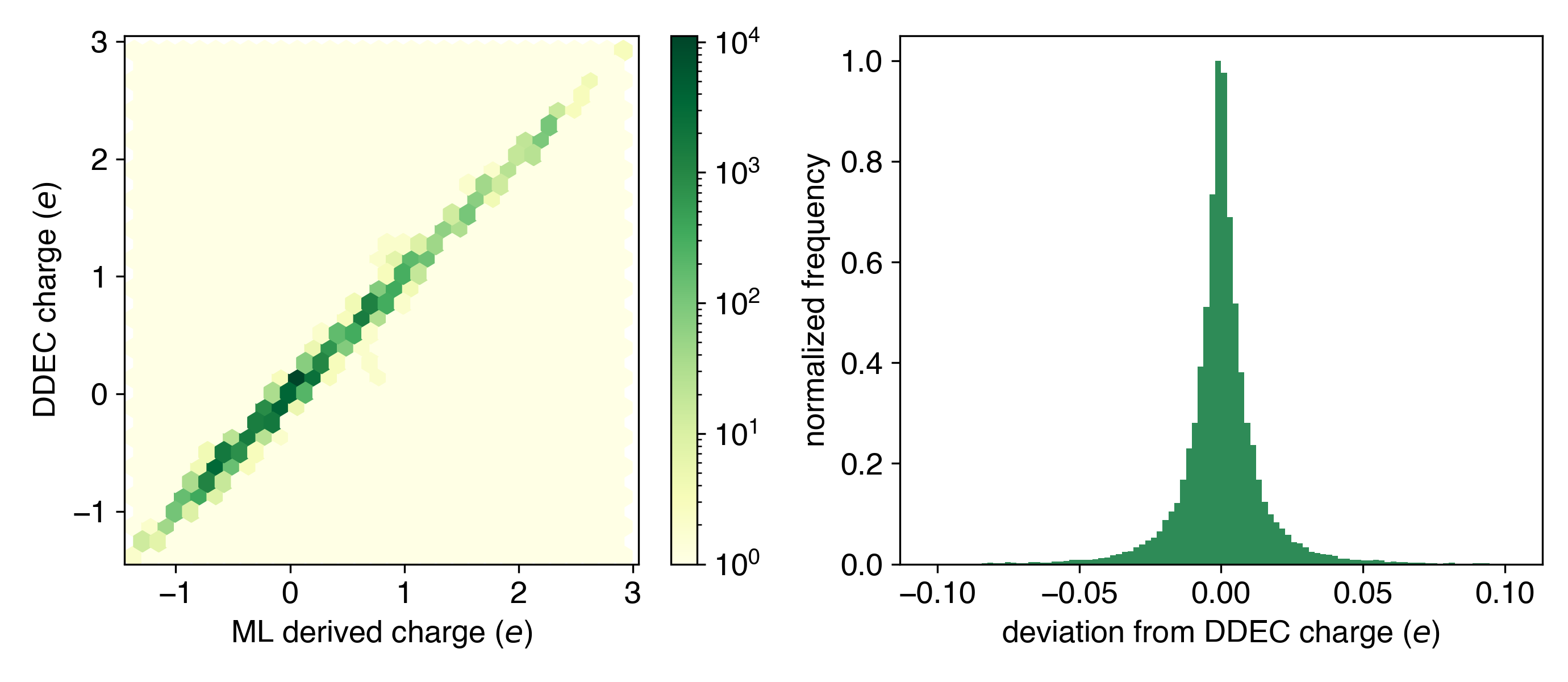}
  \caption{Predicted ML-derived partial charges vs. reference DDEC charges and a normalized histogram of the deviation of predicted ML-derived partial charges from reference DDEC charges (CoRE MOFs).}
  \label{fig:fig_2}
\end{figure}

\subsection{Descriptors}

Besides reliable input data, material representation strongly influences the performance of data-driven approach used for materials related tasks—fingerprints (so-called descriptors) serve this purpose\cite{mitchell2014machine,ghiringhelli2015big,seko2017representation,ramprasad2017machine}. There is no universal method for choosing an appropriate set of descriptors. Numerous types of descriptors varying levels of complexity have been proposed to represent materials\cite{ramprasad2017machine,schmidt2019recent}. Relevant descriptors should be simple to extract themselves and have as low dimensionality as possible. Taking into account the nature of the target property (partial atomic charge), the two types of descriptors are chosen for subsequent implementation/training of ML models (the full list of used descriptors is provided in the Supporting Information):

\begin{itemize}
  \item Intrinsic elemental properties of the corresponding site, such as atomic number, covalent radius, melting temperature.
  \item Structural descriptors of the site which characterize its local environment, such as the number of first nearest neighbors, Voronoi indices.
\end{itemize}

Thus, using the locality approximation, we determine partial atomic charge exclusively as a function of the chemical nature of the corresponding atom (through its physicochemical properties) and its local environment (depending on atoms within the cut-off radius of the considered one). This approximation enables one to predict atomic charges for different atoms independently of one another and makes the presented approach linearly scaled with a number of atoms in a unit cell in contrast to the reference DFT method.

Obviously, the presented ML scheme results in non-conserving atomic charges, since they are generated independently. To maintain the neutrality of a unit cell, we corrected the obtained charges as follows:

\[ q_{i}^{cor} = q_{i} - \frac{1}{n} \sum_{j} q_{j} = q_{i} - \frac{\Delta q}{n} \]

where \(q_{i}\) is a partial charge of \textit{i}-th site derived from the ML algorithm, \(\Delta q\) is an algebraic sum of all ML-derived charges per unit cell, \textit{n} is a number of atoms per unit cell. The quantity \(\frac{\Delta q}{n}\) (“charge deviation” per atom) reflects the deviation of the initial set \(\{q_{i}\}\) of ML-derived charges from the charge-neutral \(\{q_{i}^{cor}\}\) one.

\subsection{ML model training}

To establish relations between the suggested descriptors and the target property (partial charge), we have used gradient boosting decision trees (GBDT) method\cite{chen2016xgboost}. In addition to high accuracy, this algorithm provides feature importance, i.e., it is possible to identify the contribution of individual features to model’s performance. The set of partial charges contains approximately 440 thousand points. Ten-fold cross-validation has been implemented to evaluate the generalizability of models trained on ninety percent of available CoRE MOF 2014-DDEC charges on the external test set (ten percent of the initial dataset). A hyperparameter search has been performed using Tree-of-Parzen-Estimators (TPE) algorithm\cite{bergstra2011algorithms} implemented in Hyperopt\cite{bergstra2013making} library.

\subsection{High-throughput screening}

Grand canonical Monte Carlo\cite{frenkel2002understanding} (GCMC) simulations were carried out to determine adsorption properties of carbon dioxide in two subclasses of hybrid nanoporous materials. First, simulations were performed for 10140 CoRE MOFs at 298 K and 0.2 bar to determine volumetric CO\({}_{2}\) uptake. Dispersion interactions were modeled with mixed Lennard-Jones parameters from the Universal Force Field\cite{rappe1992uff} and TraPPE force field\cite{potoff2001vapor} for MOF atoms and adsorbed molecules, respectively; the Lorentz-Berthelot combining rules were implemented with a truncated cutoff of \SI{14}{\angstrom}. All simulation cells were expanded to at least \SI{28}{\angstrom} along each axis. Coulombic interactions were modeled by using ML-derived atomic charges for MOF atoms and TraPPE charges for CO\({}_{2}\) molecule. The CO\({}_{2}\) uptakes were computed running GCMC for 5000 initialization and 5000 production cycles. Second, the Henry coefficient and the heat of adsorption at infinite dilution were calculated for 460 CURATED (Clean, Uniform, and Refined with Automatic Tracking from Experimental Database) COFs\cite{ongari2019building} from 100,000 Widom particle insertions\cite{shing1981chemical}. Blocking spheres calculated with the Zeo++ software\cite{willems2012algorithms} were used to exclude inaccessible pores in GCMC simulations. All GCMC simulations were performed in the RASPA 2.0 molecular simulation software\cite{dubbeldam2016raspa}.

\newpage
\section{Results and discussion}
\subsection{Partial charges assignment}

Figure \ref{fig:fig_2} shows the values of the predicted partial charges vs. the corresponding DDEC charges. Mean absolute deviation and root mean squared deviation on the testing dataset (ten percent of the entire set of DDEC charges) are 0.0096 and 0.0176 e, respectively. Since partial charges are not physically observable quantities, there is no universal metric for assessing the quality of models for their prediction. As for the quantitative assessment of similar models for molecules, the value of 0.01 e can be used\cite{nebgen2018transferable,verstraelen2009electronegativity}. As was shown, such a mean absolute deviation from the value of the reference method (DFT simulations) provides reliable results for predicting practically important physicochemical properties. It is worth noting that in our case, the training dataset is an order of magnitude smaller than in the study mentioned above, and additional difficulties may occur due to periodicity and large unit cell size of MOFs.

We calculated partial charges for the publicly available subset of the CoRE MOF 2019 database  (10140 ordered structures); corresponding Crystallographic Information Files (CIFs) generated with pymatgen\cite{ong2013python} library is provided in the Supporting Information. Figure S1 contains the distribution of charge deviation. The mean absolute value of charge deviation \(\frac{\Delta q}{n}\) is only 0.013 e/atom, demonstrating the almost complete self-consistency of the method. Moreover, the CoRE MOF 2019 database also includes a subset of the so-called “disordered” structures, which contains pairs of atoms with a distance less than \SI{0.1}{\angstrom}. The mean absolute value of charge deviation for this subset equals 0.029 e/atom (Figure S1). We propose that the value of charge deviation comparable with the mean absolute deviation on the testing set (0.01 e) should be considered an indicator of the ML charge assignment self-consistency.

It should be noted that one of the reasons for the divergence of DFT calculations (the reference method for DDEC charge assignment) is that the maximum limit of virtual memory has been exceeded. Figure S2 shows the distribution of structures from the CoRE MOFs 2014 database by the number of atoms per unit cell. The largest structures with assigned charges contain no more than 584 atoms, while the CoRE MOFs database contains much larger structures with up to ten thousand atoms per unit cell. These structures can be used to test the concept of extensibility, that is, the general possibility of applying (generalizing) the predictive model to larger structures as compared to those used in the training dataset. The \textit{extensibility} of data-driven models for charge assignment was previously confirmed for molecular systems\cite{nebgen2018transferable,sifain2018discovering}. The use of structures with a relatively small unit cell as training samples is especially critical for the case of MOFs (three-dimensional periodic structures containing up to several thousand atoms) since the reference charge assignment method (DFT) scales as \(\sim N^{3}\). At the same time, the proposed method is based on using only local atomic features. Thus, the training of the model and the calculation of charges scale as \( \sim N^{1} \).

\begin{figure}[h]
  \centering
  \includegraphics[height=7cm]{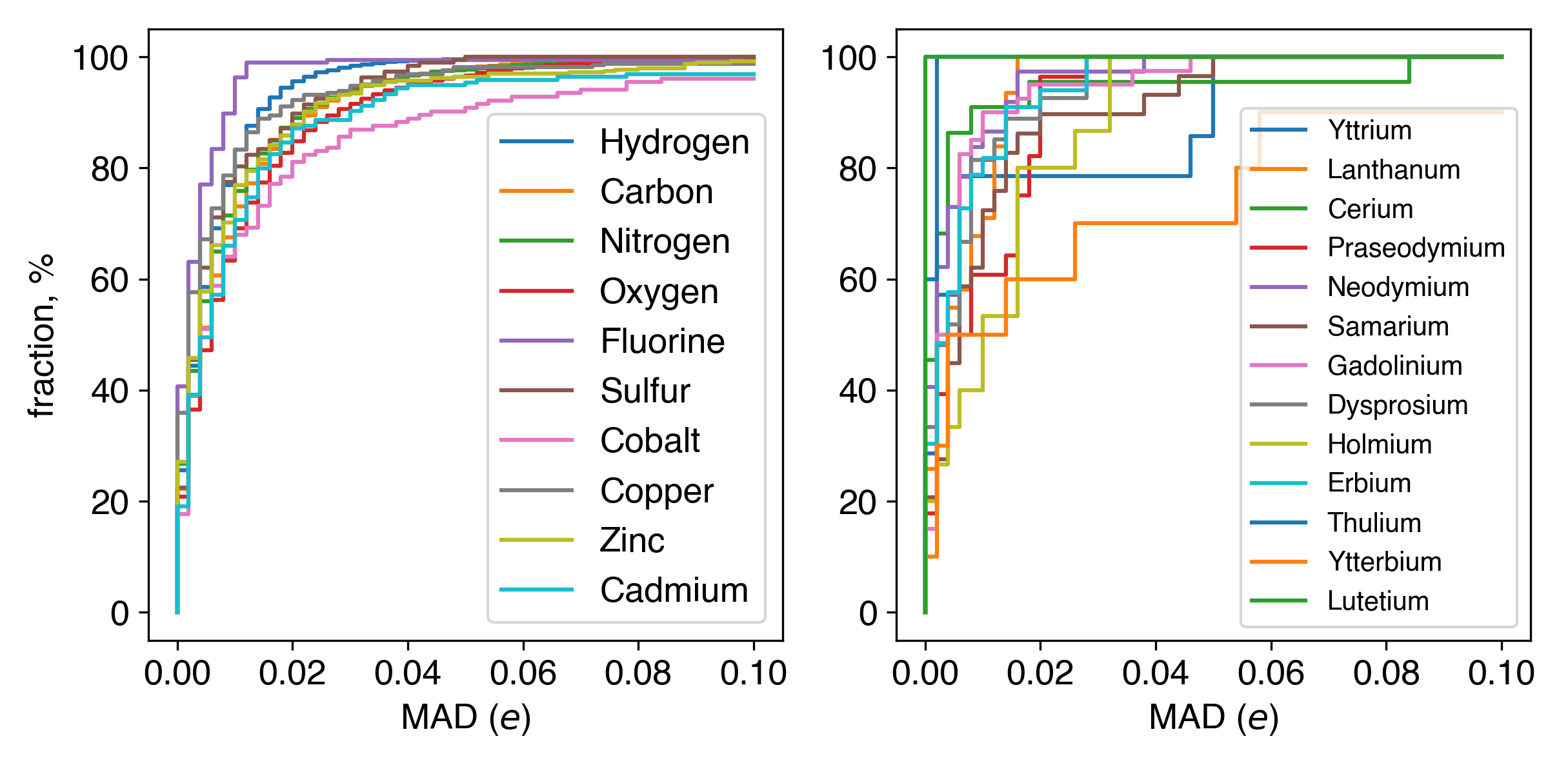}
  \caption{Cumulative fraction of samples as a function of the mean absolute deviation (MAD) for the ten most numerous elements and rare-earth elements.}
  \label{fig:fig_3}
\end{figure}

\begin{figure}[h]
  \centering
  \includegraphics[height=5cm]{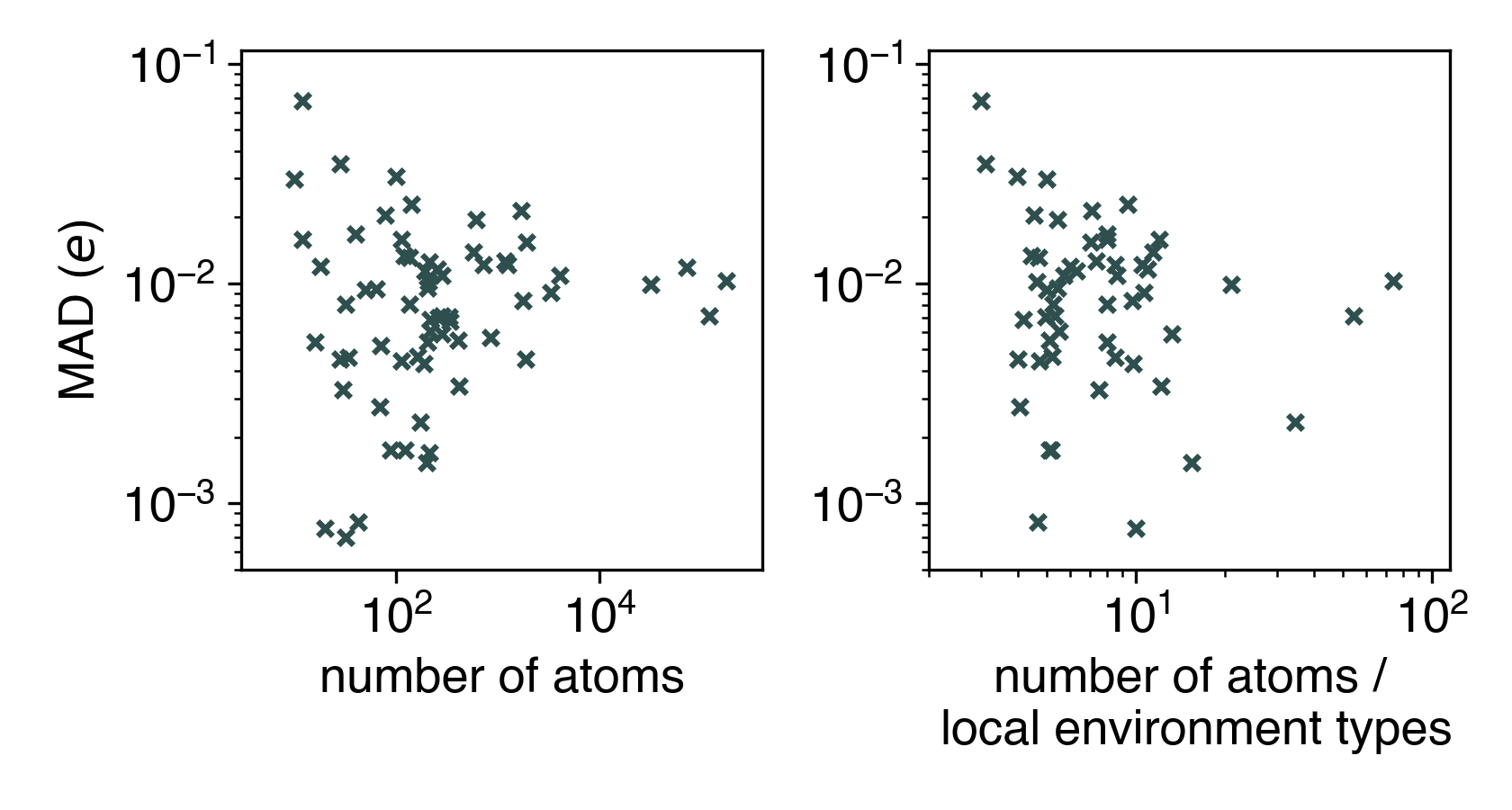}
  \caption{The mean absolute deviations (MAD) of partial charge predictions as a function of number of atoms and number of atoms per number of local environment types. Each point corresponds to a distinct chemical element.}
  \label{fig:fig_4}
\end{figure}

For a more detailed consideration of the performance of our models, we have examined its effectiveness in predicting charges for distinct chemical elements. The corresponding cumulative curves for the ten most numerous and rare-earth elements are shown in Figure \ref{fig:fig_3}. The most general trend that can be identified is that the prediction efficiency is higher for the most numerous elements. However, when considering a similar relationship for the whole periodic system of elements (Figure \ref{fig:fig_4}), it is not observed. As it may seem, it goes against the general feature of the data-dependent algorithms, “more data is better,” which is particularly valid for materials science-related applications\cite{zhang2018strategy}. Normalization to the number of different types of the local environment also did not lead to a pronounced relationship between the size of the dataset for individual chemical elements and the corresponding mean absolute deviations. The most obvious explanation is that the model’s performance is only partially determined by the nature of the chemical element and its nearest coordination sphere (the criterion by which structural types were distinguished in the original study\cite{nazarian2016comprehensive}). Thus, for Zn atoms coordinated to six O atoms, the charge value is in the range 1.18-1.85\cite{nazarian2016comprehensive}, i.e., other features influence the partial charge value as well.

\begin{figure}[h]
  \centering
  \includegraphics[height=8cm]{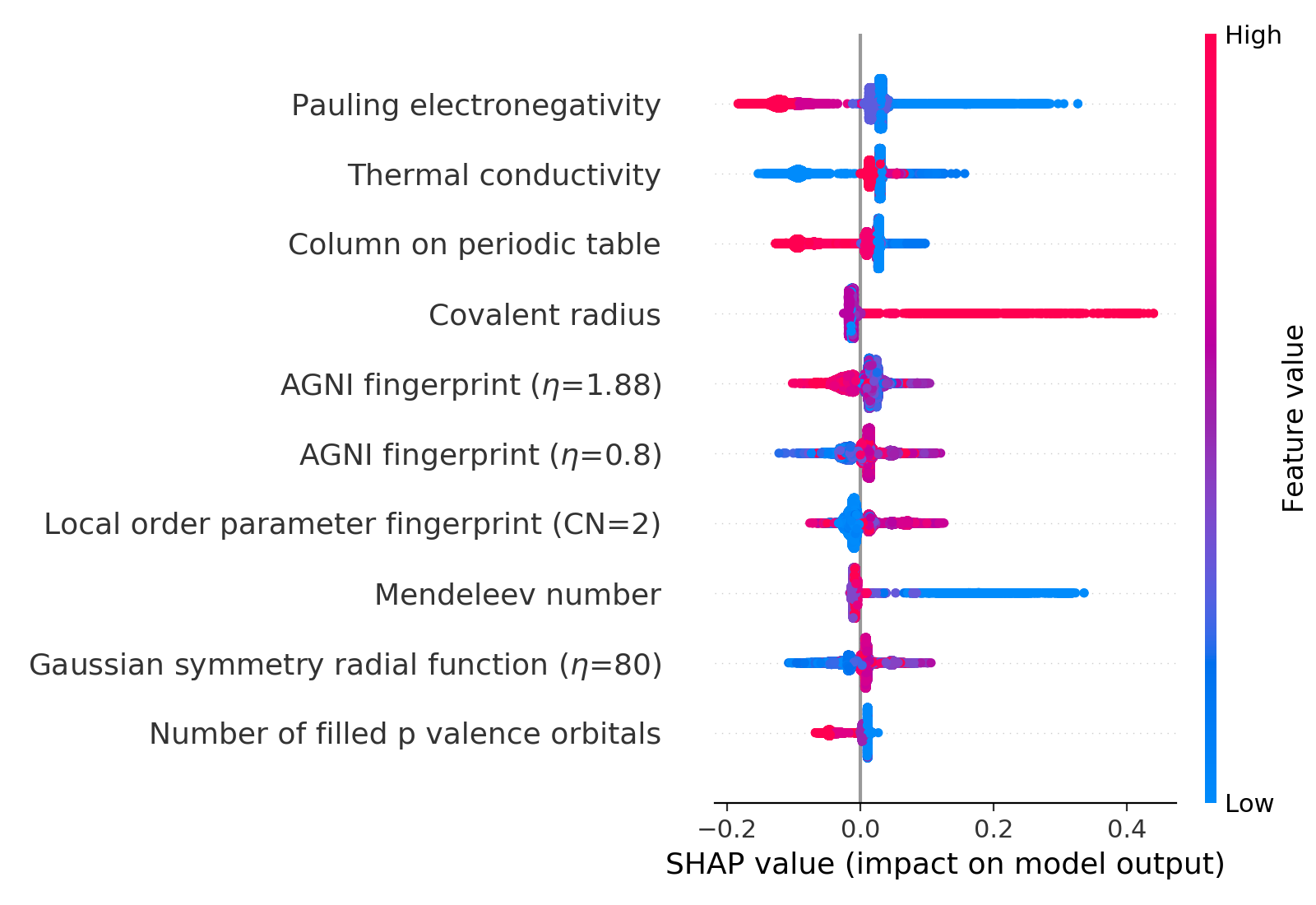}
  \caption{SHAP summary plot of the ten most valuable features of the XGBoost model for partial charge prediction. Dots are colored by the feature’s value for each atom in the test set and piled up vertically to show density.}
  \label{fig:fig_5}
\end{figure}

In order to get insights on which of the used features make the most significant contribution to the predictions, we use the SHapley Additive exPlanations (SHAP) approach implemented for tree ensemble methods\cite{lundberg2018consistent}. The SHAP summary plots for ten most-valuable features (with the highest global impact \(\sum_{n=1}^{N} {|\phi|}_{i}^{j}\), where \({|\phi|}_{i}^{(j)}\) are SHAP values) are presented in Figure \ref{fig:fig_5}. Among these features, one should first highlight physicochemical elemental properties. Not surprisingly, electronegativity is the most important feature affecting the value of partial charges. Among other elemental properties, one should also highlight the covalent radius, large values of which contribute to a positive charge (e.g., for rare-earth elements) and Mendeleev number (hydrogen stands out, for which a low value leads to an increase in positive charge).

Thus, 4 out of 10 most valuable features characterize the local environment of the atom from different perspectives:

\begin{itemize}
  \item The local order parameter\cite{zimmermann2017assessing} is a structural parameter with Voronoi-tessellation based neighbor finding. This particular fingerprint represents how consistent the local environment is with a coordination number equal to two.
  \item Adaptive Generalizable Neighbourhood Informed (AGNI) fingerprint\cite{botu2015learning,botu2015adaptive} is a fingerprint based on integrating the distances product of the radial distribution function with a Gaussian window function. The parameter \( \eta \) governs the extent of coordination around a considered atom.
  \item Gaussian symmetry radial function\cite{khorshidi2016amp,behler2011atom}, where \( \eta \) is a radial function parameter.
\end{itemize}

The above-mentioned structural descriptors can be considered as the development of the connectivity-based atom contribution (CBAC) approach. This method reduces the diversity of local environments to a very limited number (up to a hundred in the case of structurally diverse elements, such as Zn). Thus, it completely ignores the mutual position of atoms in the first coordination sphere, as well as the distance to the central atom. These factors also have a significant impact on the partial charge, whereas the structural descriptors make it possible to take into account minor differences between local environments that are identical in terms of bonding connectivity.

\begin{figure}[h]
  \centering
  \includegraphics[height=8cm]{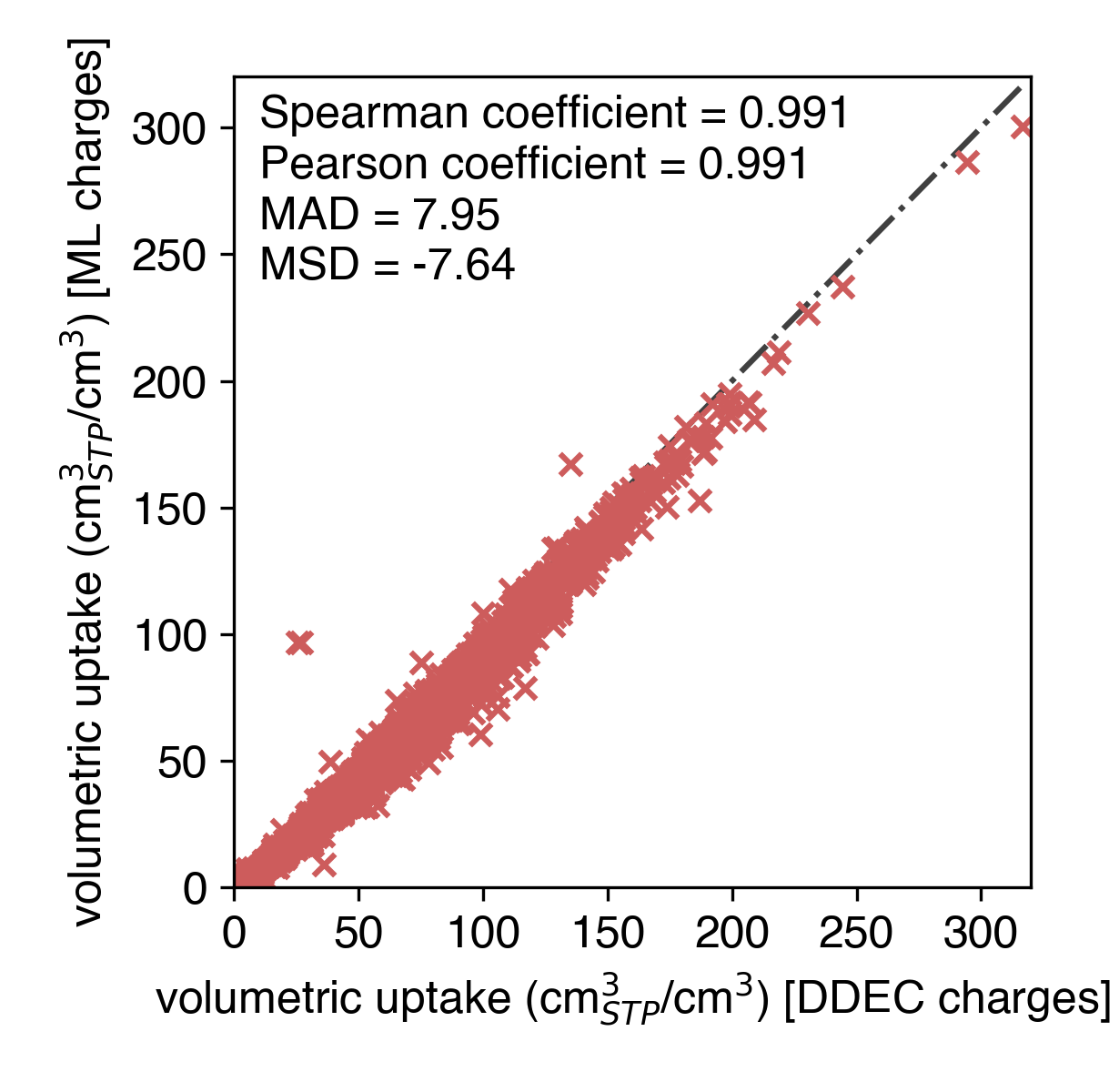}
  \caption{Comparison of the carbon dioxide volumetric uptake calculated by means of DDEC (as a reference set) and ML-derived charges (CoRE MOFs).}
  \label{fig:fig_6}
\end{figure}

In most cases, ML models in materials science are applied to predict physicochemical properties characterizing compounds from “macroscopical” perspective, such as formation energy, bandgap, and elastic moduli. In contrast, microscopic features such as atomic energies, forces, and partial charges are not the ultimate goal of the computational analysis; these quantities are necessary for building semi-empirical potential models. In particular, partial charges generated by the proposed data-driven method can be used for high-throughput computational screening of MOFs. To prove the viability of the ML charge assignment, we have addressed the challenge of carbon dioxide capture\cite{yu2017co2,trickett2017chemistry}. Environmental conditions (298 K, 0.2 bar) were chosen to directly compare the results with the data provided in the benchmark study\cite{ongari2018evaluating}, where CoRE MOF 2014-DDEC charges were used as a reference as well. Figure \ref{fig:fig_6} demonstrates the values of CO\({}_{2}\) volumetric uptake in CoRE MOFs obtained by using DDEC- and ML-derived charges. The mean absolute deviation of 7.95 \( {cm}_{STP}^{3} / {cm}^{3} \) is significantly less in comparison with any charge equilibration methods\cite{ongari2018evaluating}. Moreover, the best of them (MEPO-Qeq method) has failed to assign charges to 1566 structures, i.e., 67 percent of considered CoRE MOFs, due to its poor transferability to new types of local atomic environments.

\subsection{Transferability of ML charge assignment}

\begin{figure}[h]
  \centering
  \includegraphics[height=7cm]{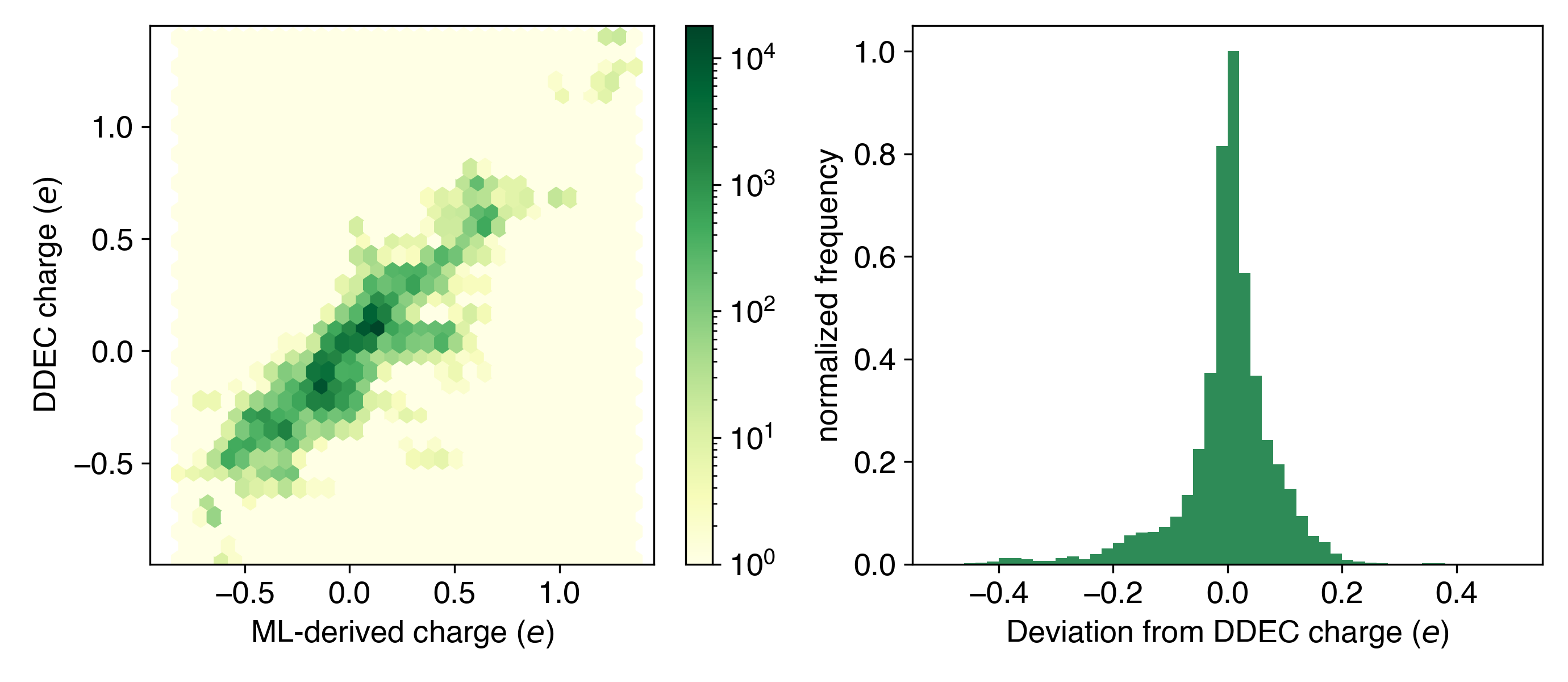}
  \caption{Predicted ML-derived partial charges vs. reference DDEC charges and a normalized histogram of the deviation of predicted ML-derived partial charges from reference DDEC charges (CURATED COFs).}
  \label{fig:fig_7}
\end{figure}

The nanoporous materials genome is not limited to MOFs. For instance, covalent organic frameworks\cite{feng2012covalent,diercks2017atom,lohse2018covalent} (COFs) are formed by directly connected ligands in contrast to MOFs, where metal ions are coordinated to nonbonded ligands. The high-throughput screening seems to be an optimal strategy to identify the most promising COFs for absorbance-related applications due to their structural diversity\cite{martin2014silico,martin2014silico2,mercado2018silico}. At the same time, the absence of high-quality DFT-derived charges for modeling COFs is even more critical than in the case of MOFs. Thus, the most extensive set\cite{ongari2019building} contains partial charges for nearly five hundreds of structures. To demonstrate the transferability of the presented model (originally intended to predict MOFs partial charges) to different subclasses of porous materials, we have evaluated charges for a set of 460 experimentally reported COFs\cite{ongari2019building}. ML-derived charges are compared with previously calculated DDEC charges in Figure \ref{fig:fig_7}. The mean absolute deviation equals 0.05 e.

\begin{figure}[h]
  \centering
  \includegraphics[height=7cm]{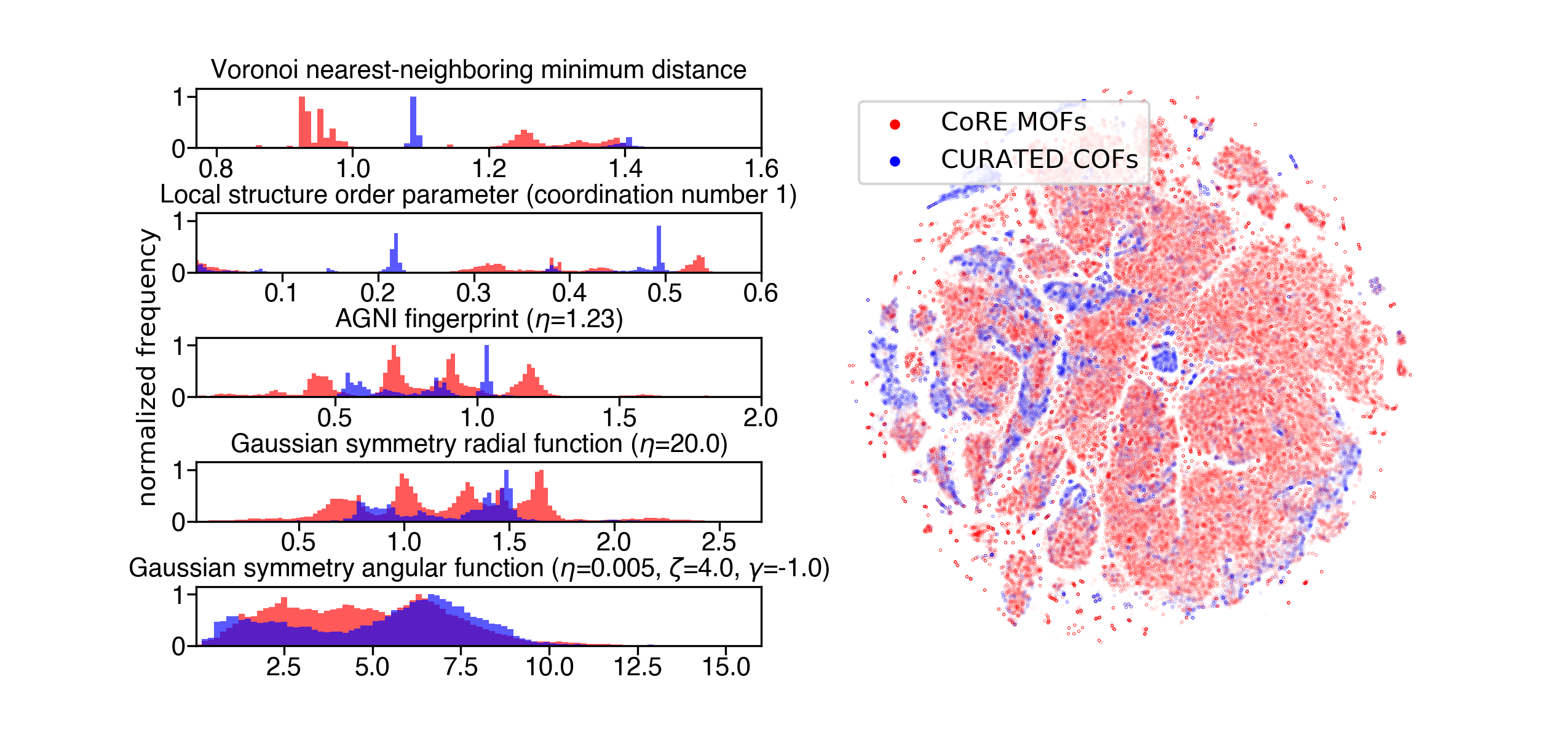}
  \caption{Distributions of several illustrative ML descriptors used to represent the local environment of MOF and COF atoms. Two-dimensional projections of all structural descriptors (Table S2) using the t-distributed stochastic neighbor embedding algorithm.}
  \label{fig:fig_8}
\end{figure}

On the one hand, the value is significantly larger than the corresponding mean absolute deviation for CoRE MOFs. The accuracy of the ML model (among other things) is determined by its applicability domain. Obviously, local environments of atoms in COFs significantly differ from that of typical for MOFs, at least because most of the considered COFs are layered two-dimensional structures. To demonstrate the discrepancy between local atomic environments in MOFs and COFs, we provide the distributions of some ML descriptors in Figure \ref{fig:fig_8}. In addition, we have also generated two-dimensional projections of all structural descriptors using the t-distributed stochastic neighbor embedding\cite{maaten2008visualizing} (t-SNE) algorithm. MOFs and COFs subdomains of feature space look weakly interconnected (Figure \ref{fig:fig_8}), which explains the significantly increased error of our model (trained on MOFs) for COFs. It should be emphasized that there was no predominant knowledge about the point belonging to a particular environment type (MOF or COF) during the t-SNE training. On the other hand, the reported value is still significantly less than the mean absolute deviation, which is typical for Qeq schemes (0.1 e) in the case of MOFs. A recently published study\cite{deeg2020silico} provides an opportunity to compare the performance of the ML approach and equilibration methods directly on two sets of 98 COFs. The presented scheme is characterized by the performance (the root mean squared deviation of 0.085 e) similar to the Qeq method (0.092 e) and its advanced modification driven by genetic algorithm (0.075 e).

\begin{figure}[h]
  \centering
  \includegraphics[height=7cm]{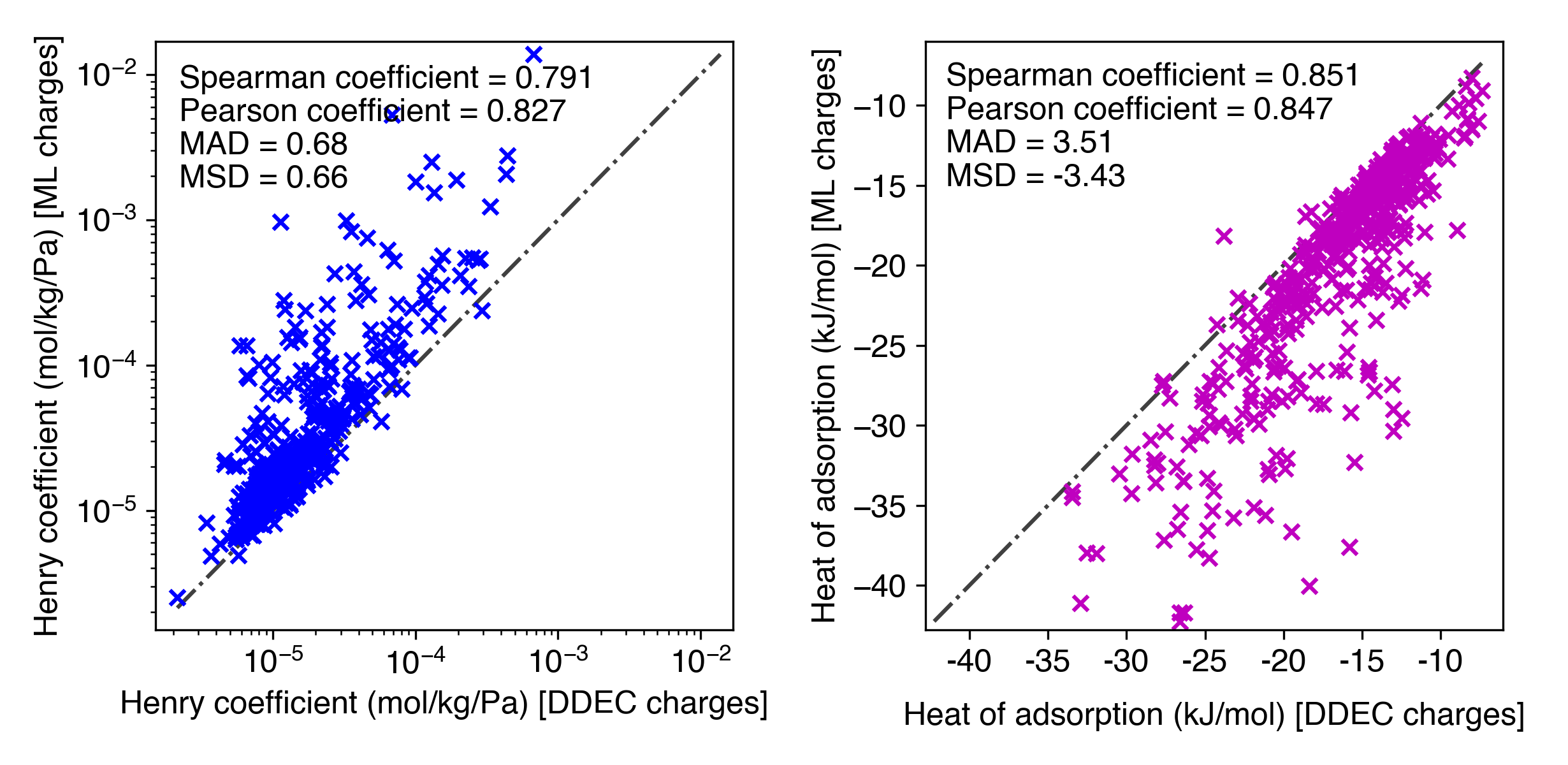}
  \caption{Comparison of the Henry coefficient and the heat of adsorption calculated by means of DDEC (as a reference set) and ML-derived charges (CURATED COFs).}
  \label{fig:fig_9}
\end{figure}

To assess the impact of ML-derived charges on the adsorption properties of carbon dioxide, we have evaluated the Henry coefficient and the heat of adsorption in the infinite dilution regime for 460 COFs and compared the results with the values obtained using DDEC charges (Figure \ref{fig:fig_9}). As can be seen from the mean absolute and signed (MSD) deviations, the Henry coefficient and heat of adsorption are systematically overestimated so that the corresponding mean absolute deviations may be reduced to 0.46 and 2.08 by multiplication by a correction factor of 0.65 and 0.86, respectively. However, high-throughput computational screenings are mainly aimed at ranking promising candidates for a specific application. From this perspective, nonparametric measures of rank correlation seem to be more relevant metrics for our case. Spearman’s rank correlation coefficient between Henry’s constant values obtained through the use of high-quality DDEC charges and ML-derived charges equals 0.79; the same value for the heat of adsorption equals 0.85. The Henry coefficient is highly correlated\cite{ongari2019building} with parasitic energy\cite{huck2014evaluating} defined as electric energy required for flue gas separation and subsequent CO\({}_{2}\) storage. Thus, high Spearman’s rank correlation coefficient for the Henry’s constant proves that ML-derived charges seem to be a reasonable choice for initial large-scale screening of porous materials.

\section{Conclusions}

The presented ML approach does not require any additional parameters and can be used out of the box, which frees chemists from the need to select a method/parameters, and allows focusing on modeling the adsorption properties of nanoporous materials. Moreover, ML-derived charges make it possible to reproduce reference data much more accurately in comparison with computationally low-cost equilibration methods.
Since the data-driven approach relates to the use of continuous local features, it allows assigning charges to structures previously inaccessible for first principal calculations, i.e., containing heavy elements (such as europium or terbium) or thousands of atoms per unit cell. We provide a publicly available graphical interface to the developed model and ML-derived charges for 10140 CoRE MOF structures to facilitate high-throughput computational screening.

The graphical interface is available on \url{https://mof.scidatasoft.com/api/ui/#!/predict/app_charges}.

\section{Acknowledgment}

The research is carried out using the equipment of the shared research facilities of HPC computing resources at Lomonosov Moscow State University.

\end{document}